\documentclass{jpsj2}
\title{Adiabatic Phase Diagram of an Ultracold Atomic Fermi Gas \\
with a Feshbach Resonance}

\author{Shohei \textsc{Watabe}$^{1}$
, Tetsuro \textsc{Nikuni}$^{2}$
, Nicolai \textsc{Nygaard}$^{3}$
, James E. \textsc{Williams}$^{4}$
,\\ 
and Charles W. \textsc{Clark}$^{4}$
}

\inst{$^{1}$Department of Physics, 
Graduate School of Science, 
University of Tokyo, 
7-3-1 Hongo, Bunkyo-ku, Tokyo, Japan 
113-0033 \\
$^{2}$Department of Physics, Faculty of Science, 
Tokyo University of Science, 
1-3 Kagurazaka, Shinjuku-ku, Tokyo, 
Japan, 162-8601 \\
$^{3}$Lundbeck Foundation Theoretical Center for Quantum System Research,
Department of Physics and Astronomy, 
University of Aarhus, 
DK-800 {\AA}rhus C, Denmark \\
$^{4}$
Electron and Optical Physics Division, 
National Institute of Standards 
and Technology, 
Gaithersburg, MD 20899-8410, USA
}

\abst{
We determine the adiabatic phase diagram 
of a resonantly-coupled system of Fermi atoms 
and Bose molecules confined in the harmonic trap 
by using the local density approximation. 
The adiabatic phase diagram shows 
the fermionic condensate fraction 
composed of condensed molecules and Cooper pair atoms. 
The key idea of our work is conservation of entropy 
through the adiabatic process, 
extending the study of 
Williams {\it et al}.
[Williams {\it et al}., New J. Phys. {\bf 6}, 123 (2004)]
for an ideal gas mixture to include the resonant interaction 
in a mean-field theory. 
We also calculate the molecular conversion efficiency 
as a function of initial temperature. 
Our work helps to understand 
recent experiments on the BCS-BEC crossover, 
in terms of the initial temperature 
measured before a sweep of the magnetic field.
}

\kword{degenerate Fermi gas, Feshbach resonance, BCS-BEC crossover,  adiabatic sweep, conversion efficiency}

\begin{document}
\maketitle

\section{Introduction}
Degenerate Fermi gases with controllable interaction 
have been realized in recent experiments 
by making use of a Feshbach resonance. 
Much of the current study of these systems focuses on the crossover 
between a Bardeen-Cooper-Schrieffer (BCS) superfluid of fermionic atoms 
and Bose-Einstein condensation (BEC) 
of diatomic molecules~\cite{regal,bartenstein,zwierlein,bourdel}. 
In these systems, the position of a molecular bound state 
can be tuned relative to the scattering continuum by 
varying an applied magnetic field. 
By ramping the magnetic field across a Feshbach resonance, 
one can continuously transform 
a fermionic atomic gas to a bosonic molecular gas, 
both above and below the superfluid transition temperature. 

Several theoretical papers reported the phase diagrams of the 
resonantly-coupled Fermi-Bose mixture gas~\cite{ohashi,falco,diener}. 
In order to compare these works with the experimental results 
by Regal {\it et al}.~\cite{regal} 
and Zwierlein {\it et al}.~\cite{zwierlein}, 
one has to understand how the system traverses 
the phase diagram as the resonance energy is varied. 
However, the experiments only provided a measure of the
{\textit{initial}} temperature, $T_{i}$, before the magnetic field sweep, thus
complicating the attempts to relate the data to theoretical models.
In the experiments, the sweep is slow 
so that atoms and molecules can move and collide 
sufficiently in the trap~\cite{zwierlein}, 
allowing for relaxation to thermal and chemical equilibrium 
during the sweep. Assuming the sweep to be adiabatic , 
Williams {\it et al}.~\cite{williams} calculated  
the phase diagram for the ideal gas mixture of fermionic atoms 
and bosonic molecules as a function of the resonance energy and $T_{i}$. 
The key idea of this adiabatic process is 
the conservation of entropy during the sweep: the system follows a path of
constant entropy through the phase diagram, and the entropy is set
by the initial preparation of the sample, {\textit{i.e.}} the initial temperature. 
Using the reverse logic, 
Chen  {\it et al}.~\cite{chen} gave the idea of thermometry, 
by which the real temperature  
of the system after an adiabatic sweep may be 
deducted from the initial temperature before the sweep. 
Based on this finite temperature formalism, 
they compared the experimental phase diagram
with their theoretical boundary 
between the normal phase and 
the superfluid phase~\cite{chen2}. Similar ideas were proposed by Hu {\textit{et al.}}~\cite{Hu2006} and Carr {\textit{et al.}}~\cite{carr2004g}. 

In this paper, 
we extend the work by 
Williams {\it et al}.~\cite{williams} 
for an ideal gas mixture of fermionic atoms
and bosonic molecules to include the resonant interaction 
using mean-field theory. 
Our mean-field treatment allows for the description of 
the BCS superfluidity due to the interaction, 
which was neglected in ref. 8. 
In order to introduce the effect of the harmonic trap, 
we apply the local density approximation (LDA).  

Calculating the entropy as a function of the temperature 
and the detuning, 
we show the paths of constant entropy 
traversed in the conventional phase diagrams 
as the detuning is varied adiabatically. 
On the basis of the adiabatic path, 
we determine the adiabatic phase diagrams 
against the detuning and entropy, 
labeled by the initial temperature of the Fermi
gas measured before a sweep of the magnetic field. 
We note that the adiabatic phase diagram of ref. 10 
plots the superfluid fraction, 
whose phase boundary 
has a good agreement with the experiment in ref. 1, 
while the experimental phase diagrams~\cite{regal,zwierlein} 
show the condensate fraction 
which is zero momentum molecules 
projected through the fast sweep, 
assuming that the fraction 
has an information 
of zero momentum molecules and 
zero center-of-mass momentum Cooper pairs. 
In this paper, we plot the phase diagrams for 
the condensate fraction 
that is composed of 
zero momentum molecules and 
zero momentum Cooper pairs.  

We are also interested in 
the production efficiency of molecules 
during the adiabatic sweep of the magnetic field. 
The conversion efficiency is found experimentally to be a monotonic function 
of the initial peak phase space density, 
when the magnetic field is swept adiabatically~\cite{hodby}. 
A recent paper by 
Williams {\it et al}.~\cite{williams-conversion} 
discussed the mechanism of molecular formation, in the case where three-body recombination may be neglected, 
and derived an expression 
for the molecular conversion efficiency 
in terms of the initial peak phase-space density. 
Their result agrees with the experimental data~\cite{hodby}
although they treat a non-interacting quantum gas. 
We extend the calculation of the molecular conversion efficiency in ref. 14 
to include the mean-field effect. 

Our study gives an intuitive understanding 
of recent experiments on the BCS-BEC crossover 
using the adiabatic sweep process.

\section{Equilibrium Theory}
Our model is the coupled boson-fermion  Hamiltonian
with a two-component Fermi gas~\cite{falco,matt,kawaguchi}; 
\begin{eqnarray}
\hat{H}&=&\int d {\bf r}
[
\hat{\psi}^{\dag}_{\sigma}({\bf r})
H_{a}({\bf r})
\hat{\psi}^{}_{\sigma}({\bf r})
+
\hat{\phi}^{\dag}_{}({\bf r})
H_{m}({\bf r})
\hat{\phi}^{}_{}({\bf r})
]
\nonumber
\\
&&
+
\kappa\int d {\bf r}
[
\hat{\phi}^{\dag}_{}({\bf r})
\hat{\psi}^{}_{\uparrow}({\bf r})
\hat{\psi}^{}_{\downarrow}({\bf r})
+
\hat{\psi}^{\dag}_{\downarrow}({\bf r})
\hat{\psi}^{\dag}_{\uparrow}({\bf r})
\hat{\phi}^{}_{}({\bf r})
], 
\label{hamiltonian}
\end{eqnarray}
where repeated indices $\sigma$ imply a sum over 
pseudospin states $\sigma = \{\uparrow, \downarrow\}$. 
The field operators of atoms and molecules are respectively 
denoted as 
$\hat{\psi}^{}_{\sigma}({\bf r})$
and
$\hat{\phi}^{}_{}({\bf r})$, 
which 
obey Fermi and Bose commutation relations.
The single particle Hamiltonians for the atoms and molecules 
are given by 
$H_{a}({\bf r}) = - (\hbar^{2}/2m)\nabla^{2} + V_{a} ({\bf r})$
and
$H_{m}({\bf r}) = - (\hbar^{2}/2M)\nabla^{2} + V_{m} ({\bf r})
+ \delta $
, respectively, 
where $m$ is the atomic mass 
and $M = 2m$ is the molecular mass. 
We introduce the harmonic trap 
in which the each pseudo-spin Fermi atoms 
feel the same potential, given by 
$V_{a} ({\bf r})  =  
m \omega_{0}^{2}r^{2}/2$. 
On the other hand, we assume that 
molecules feel the following potential:
$V_{m} ({\bf r})  =  M\omega_{0}^{2}r^{2}/2$. 
This assumption is valid for 
the current experiments in optical dipole traps~\cite{zwierlein2}.  
Two fermionic atoms are coupled to a bosonic molecule
through the resonant interaction with the coupling constant $\kappa$.
The detuning $\delta(B)$
 can be tuned by adjusting an external magnetic field $B$, since the bare atoms and molecules have different magnetic moments, and thus experience a relative linear Zeeman shift. As usual we take the zero of energy to be the energy of two separated atoms at rest at each value of $B$. 
We neglect the non-resonant 
atom-atom interaction, 
since the effective  interaction for 
two atoms is dominated by the resonant contribution 
$-\kappa^{2}/ \delta$ 
near the resonance.

In order to impose the constraint that the number of particles 
be conserved, 
we introduce the chemical potential 
and deal with the grand canonical Hamiltonian 
$\hat{{\mathcal H}} \equiv \hat{H} - \mu \hat{N}$, 
where 
the total number operator of particles is defined as
$\hat{N} \equiv \int d{\bf r}
[
\sum\limits_{\sigma}
\hat{\psi}^{\dag}_{\sigma}({\bf r})
\hat{\psi}^{}_{\sigma}({\bf r})
+2
\hat{\phi}^{\dag}_{}({\bf r})
\hat{\phi}^{}_{}({\bf r})
]$. The factor of two in the last term reflects that each molecules 
are assembled from two atoms.
We regard the system as in chemical equilibrium 
between atoms and molecules~\cite{williams,nikuni}. 

In the superfluid phase, 
we can separate the molecular field operator into two parts: 
\begin{eqnarray}
\label{ }
\hat{\phi}({\bf r})&=& 
\langle \hat{\phi}({\bf r}) \rangle
+\tilde{\phi}({\bf r})
\nonumber
\\
&=&
\Phi({\bf r})
+ 
\tilde{\phi}({\bf r}).
\end{eqnarray} 
Here, 
$\Phi({\bf r})$ represents the molecular condensate wavefunction, 
which corresponds to the order parameter, 
and 
$\tilde{\phi}({\bf r})$ represents the non-condensate molecules. 
In the mean-field theory, 
one leaves out the Cooper pairs outside the condensate, 
and 
the Hamiltonian $\hat{\mathcal{H}}$ reduces to 
the BCS-type Hamiltonian~\cite{matt} 
\begin{eqnarray}
\hat{\mathcal{H}}&=&
\int d {\bf r}
\{
\Phi^{*}({\bf r})
[ H_{m}({\bf r})  - 2 \mu ]
\Phi({\bf r})
\}
+
\int d {\bf r}
\{
\tilde{\phi}^{\dagger}({\bf r})
[ H_{m}({\bf r})  - 2 \mu ]
\tilde{\phi}({\bf r})
\}
\\
&&
+
\int d {\bf r}
\{
\hat{\psi}^{\dag}_{\sigma}({\bf r})
[ H_{a}({\bf r}) - \mu ]
\hat{\psi}^{}_{\sigma}({\bf r})
\}
+
\kappa\int d {\bf r}
[
\Phi^{*}_{}({\bf r})
\hat{\psi}^{}_{\uparrow}({\bf r})
\hat{\psi}^{}_{\downarrow}({\bf r})
+
\hat{\psi}^{\dag}_{\downarrow}({\bf r})
\hat{\psi}^{\dag}_{\uparrow}({\bf r})
\Phi^{}_{}({\bf r})
],
\nonumber
\end{eqnarray}
where fluctuation terms like $\tilde{\phi}^{\dagger}\hat{\psi}_{\downarrow}\hat{\psi}_{\uparrow}$ have been neglected.

This type of quadratic Hamiltonian can be diagonalized 
by the standard Bogoliubov transformation. 
In order to discuss the thermodynamics of the system, 
we evaluate the grand canonical potential $\Omega$. 
The effect of a harmonic trap 
in the grand canonical potential $\Omega$ 
is included by the local density approximation~\cite{ohashi}, 
which adopts the effect of the local potential 
by replacing the chemical potential $\mu$ 
with $\mu({\bf r}) = \mu - V_{a}({\bf r})$. 

The grand canonical potential 
including the effect of the harmonic trap potential 
is 
given by 
\begin{eqnarray}
\Omega &=& 
\int d{\bf r} 
\left  (
[\delta-2\mu({\bf r})]
|\Phi({\bf r})|^{2}
+ 
\frac{1}{\beta}
\int \frac{d{\bf k}}{(2\pi^{3})}
\ln{\{1-e^{-\beta[\varepsilon_{\bf k}^{(m)}+\delta-2\mu({\bf r})]}\}}
\right .
\\
 & & 
+
\left .
\int \frac{d{\bf k}}{(2\pi^{3})}
\left \{
[\varepsilon_{\bf k}^{(a)}-\mu({\bf r})-E_{\bf k}({\bf r})]
-
\frac{2}{\beta}
\ln{[1+e^{-\beta E_{\bf k}({\bf r})}]}
\right \}
\right ) .
\end{eqnarray}
Here, 
$ \varepsilon^{\scriptsize{(m)}}
_{{\bf k}} = 
\hbar^{2} {\bf k}^{2}/2M$
is the kinetic energy of a molecule, 
and 
$ \varepsilon^{\scriptsize{(a)}}
_{{\bf k}} 
= \hbar^{2} {\bf k}^{2}/2m
$
is the kinetic energy of an atom. 
As usual $\beta=1/kT$, where $T$ is the temperature and $k$ is Boltzmann's constant. 
At this step, we have already assumed that 
the atoms and molecules are in thermal equilibrium.
In other words, the atom temperature and the molecule temperature 
are the same~\cite{williams,nikuni}.  
The local quasiparticle excitation energy is
$E_{{\bf k}} ({\bf r})= 
\sqrt{[\varepsilon_{{\bf k}}
^{(a)}-\mu ({\bf r})]^{2}
+|\Delta ({\bf r})|^{2}}$, where the local gap energy is
 $|\Delta ({\bf r})| = \kappa |\Phi ({\bf r})|$.
Within our mean-field theory, 
the excitation spectrum of bosonic molecules 
has the finite excitation gap $\delta - 2 \mu$. 
However, in the superfluid phase, 
the symmetry breaking requires a gapless spectrum 
with the Bogoliubov phonon mode. 
This can be incorporated into the theory 
by including the effect of fluctuations~\cite{ohashi4}, 
which are neglected in our mean-field theory. 
The gapless excitations are important at very low temperatures. 
We will discuss this point in {\S} 3.

The local gap equation is obtained
 from $\delta \Omega / \delta |\Phi ({\bf r})| = 0$ 
, which leads to 
\begin{equation}
\delta - 2\mu({\bf r}) =
\frac{\kappa^{2}}{2}
\int \frac{d{\bf k}}{(2\pi)^{3}}
\frac{1}{E_{{\bf k}}({\bf r})}
\tanh{\left [ \frac{\beta E_{{\bf k}}({\bf r})}{2} \right ]}
.
\label{gapeq}
\end{equation}
On the other hand, the total number of particles 
 $N_{\rm{tot}}$ follows from the relation 
$ N_{\rm{tot}} = - \partial \Omega/\partial \mu$,
which gives
\begin{eqnarray}
 N_{\rm{tot}}&=&
2\int d{\bf r}
|\Phi ({\bf r})|^{2}
+ 
2
\int \frac{d{\bf r}d {\bf k}}{(2\pi)^{3}}
f_{B}(\varepsilon_{{\bf k}}^{{\scriptsize{(m)}}};{\bf r})
\nonumber
\\
&&
+ 
\int \frac{d{\bf r} d {\bf k}}{(2\pi)^{3}}
\left[
1-\frac
{\varepsilon_{{\bf k}}
^{{\scriptsize{(a)}}}-\mu({\bf r})}
{E_{{\bf k}}({\bf r})}
+
2
\frac
{\varepsilon_{{\bf k}}
^{{\scriptsize{(a)}}}-\mu({\bf r})}
{E_{{\bf k}}({\bf r})}
f_{F}(E_{{\bf k}}^{};{\bf r})
\right]  .
\label{density}
\end{eqnarray} 
Here, the Bose and Fermi local distribution functions are 
respectively 
$f_{B}(\varepsilon_{{\bf k}}
^{\scriptsize{(m)}};{\bf r})$ 
and 
$f_{F}(E_{{\bf k}}
^{};{\bf r})$, 
which are given by 
$f_{B}
(\varepsilon_{{\bf k}}
^{\scriptsize{(m)}};{\bf r}) = 1/\{ \exp{[\beta
(\varepsilon_{{\bf k}}
^{\scriptsize{(m)}} + \delta - 2 \mu({\bf r})
)]}-1
\}$
and
$f_{F}
(E_{{\bf k}}
^{};{\bf r}) = 1/[ \exp{(\beta
E_{{\bf k}}({\bf r}))}+1
 ]$ .
 In the right-hand-side of eq. (\ref{density}), 
 the first term represents 
 the twice number of the condensed molecules $2N_{mc}$.  
 The second term represents 
 the twice number of the non-condensed molecules $2\tilde{N}_{m}$. 
The last term is 
 the number of atoms $N_{a}$.   
 Thus, eq. (\ref{density}) can be written as 
 \begin{eqnarray}
 N_{\rm tot} &=& 2N_{m} + N_{a}
 \nonumber
 \\
 &=&2N_{mc}+2\tilde{N}_{m} + N_{a}. 
 \end{eqnarray}

We also calculate the condensed pair number $N_{c}$, 
which is composed of the number of 
condensed molecules $N_{mc}$ and Cooper pairs $N_{p}$: 
\begin{eqnarray}
 N_{c} &=& N_{mc} + N_{p}
 \nonumber
 \\
 &=&
\int d{\bf r} |\Phi ({\bf r})|^{2} + 
\int d {\bf r}_{1} d {\bf r}_{2} 
|\langle \hat{\Psi}_{\uparrow} ({\bf r}_{1})
\hat{\Psi}_{\downarrow} ({\bf r}_{2})\rangle|^{2} , 
\label{Nc}
\end{eqnarray}
where the second term of eq. (\ref{Nc}) 
describing the number of the condensed Cooper pair 
is the maximum eigenvalue of two-particle density matrix~\cite{yang}. 
Within our mean-field theory, the explicit expression 
for the number of Cooper pairs $N_{p}$ is given  by~\cite{szymanska,salasnich}
\begin{eqnarray}
\label{ }
N_{p}&=&
\int \frac{d{\bf r}d{\bf k}}{(2\pi)^{3}}
\left \{
\frac{|\Delta ({\bf r})|^{2}}{4E^{2}_{{\bf k}}({\bf r})}
\tanh^{2}{\left [\frac{\beta E_{{\bf k}}({\bf r})}{2}\right ]}
\right \} .
\label{Np}
\end{eqnarray}

We regard eqs. (\ref{gapeq}) and (\ref{density}) 
as the simultaneous equations for 
the chemical potential $\mu$ and the local gap $|\Delta ({\bf r})|$, 
for a given temperature $T$ and detuning $\delta$. 
In this paper, 
we set the bare coupling strength to 
$\alpha \equiv \sqrt{\rho} \kappa = 0.4 \varepsilon_{\rm F}$, 
where $\rho$ is the peak density and $\varepsilon_{\rm F}$ is the Fermi energy 
of a pure gas of fermionic atoms, 
which corresponds to the narrow Feshbach resonance, 
which is given by 
$\rho = (2m\varepsilon_{\rm F})^{\frac{3}{2}}/(3\pi^{2}\hbar^{3})$ 
and 
$\varepsilon_{\rm F} = (3N_{\rm tot})^{\frac{1}{3}}\hbar \omega_{0}$. 
This weak coupling constant allows for 
a mean-field treatment. 
Although this value does not correspond to 
experimental situation of the broad Feshbach resonance, 
discussed in ref. 23 
(see also discussion in {\S} 3), 
we do not expect any qualitative differences in equilibrium phase diagrams 
between narrow and broad Feshbach resonances. 
To avoid the divergence of the integral in eq. (\ref{gapeq}), 
we impose the Gaussian cut-off; 
$\exp{ \{ - [ \varepsilon/(2\varepsilon_{\rm F})]^{2} \} }$, 
whose cutoff energy scale is chosen to be the Fermi energy~\cite{ohashi}.

\section{Adiabatic Phase Diagram} 
In this section, we discuss the adiabatic phase diagram. 
We first briefly review the experimental procedures 
of refs. 1 and 3. 
In these experiments, the ultracold two-component Fermi gas 
is initially prepared at a magnetic field
far detuned from the resonance position, corresponding to a pure atomic gas. 
In this weakly interacting regime the temperature of the Fermi gas is measured 
using time-of-flight imaging.
The magnetic field is then slowly lowered to the vicinity of the resonance
 to allow the atoms and molecules sufficient time to move and collide 
in the trap. 
This process may satisfy the condition 
that the gas be able to collisionally relax to equilibrium~\cite{williams}. 
We regard this process as an adiabatic quasistatic process, 
because the magnetic field is varied while keeping the thermal and chemical 
equilibrium from moment to moment 
without exchanging heat or particles 
with the environment. 
The system will thus follow a path of constant entropy in the phase diagram. 

The total entropy $S_{\rm tot}$ 
is the sum of the atom entropy $S_{a}$ and 
the molecule entropy $S_{m}$.
These are given by 
\begin{eqnarray}
S_{a}&=&
2 k_{B}
\int \frac{d{\bf r}d{\bf k}}{(2\pi)^{3}}
\left\{
\frac{\beta E_{{\bf k}}({\bf r})}
{ e^{\beta E_{{\bf k}}({\bf r})}+1}
+ \ln{[1+e^{-\beta E_{{\bf k}}({\bf r})}]}
\right\},
\label{Sa}
\end{eqnarray}
and
\begin{eqnarray}
S_{m}&=&
k_{B}
\int \frac{d{\bf r}d{\bf k}}{(2\pi)^{3}}
\left(
\frac{\beta [
\varepsilon_{{\bf k}}
^{\scriptsize{(M)}}
+\delta - 2 \mu({\bf r})]}
{ e^{\beta
 [\varepsilon_{{\bf k}}
^{\scriptsize{(M)}} 
+\delta - 2 \mu({\bf r})]}-1}
- \ln{\{1-e^{-\beta 
[\varepsilon_{{\bf k}}
^{\scriptsize{(M)}} + \delta - 2 \mu({\bf r})]}\}}
\right).
\label{Sm}
\end{eqnarray}

\begin{figure}
\begin{center}
\begin{minipage}{0.4\hsize}
\begin{center}
\includegraphics[width=6cm,height=6cm,keepaspectratio,clip]{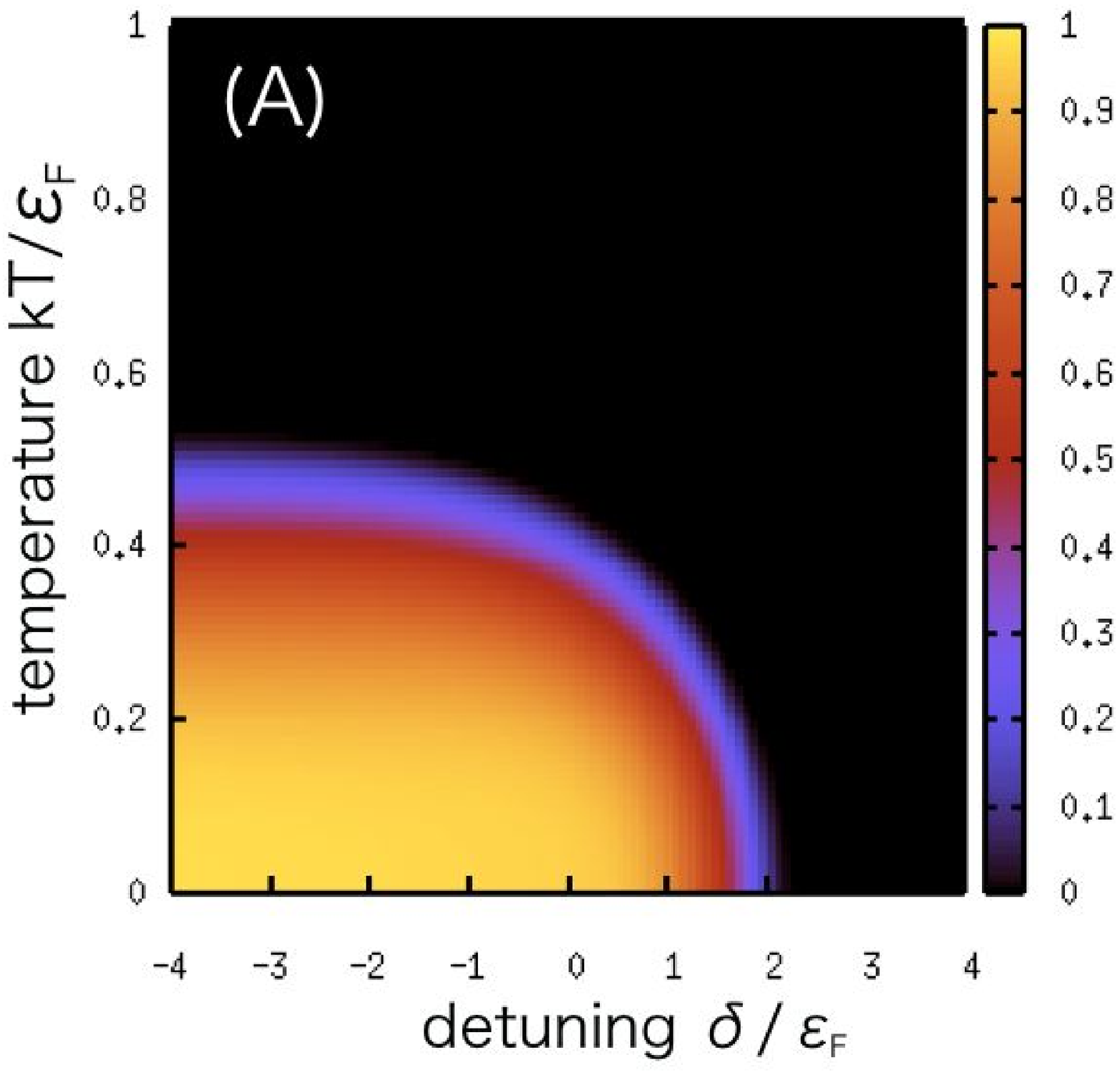}
\end{center}
\begin{center}
\includegraphics[width=6cm,height=6cm,keepaspectratio,clip]{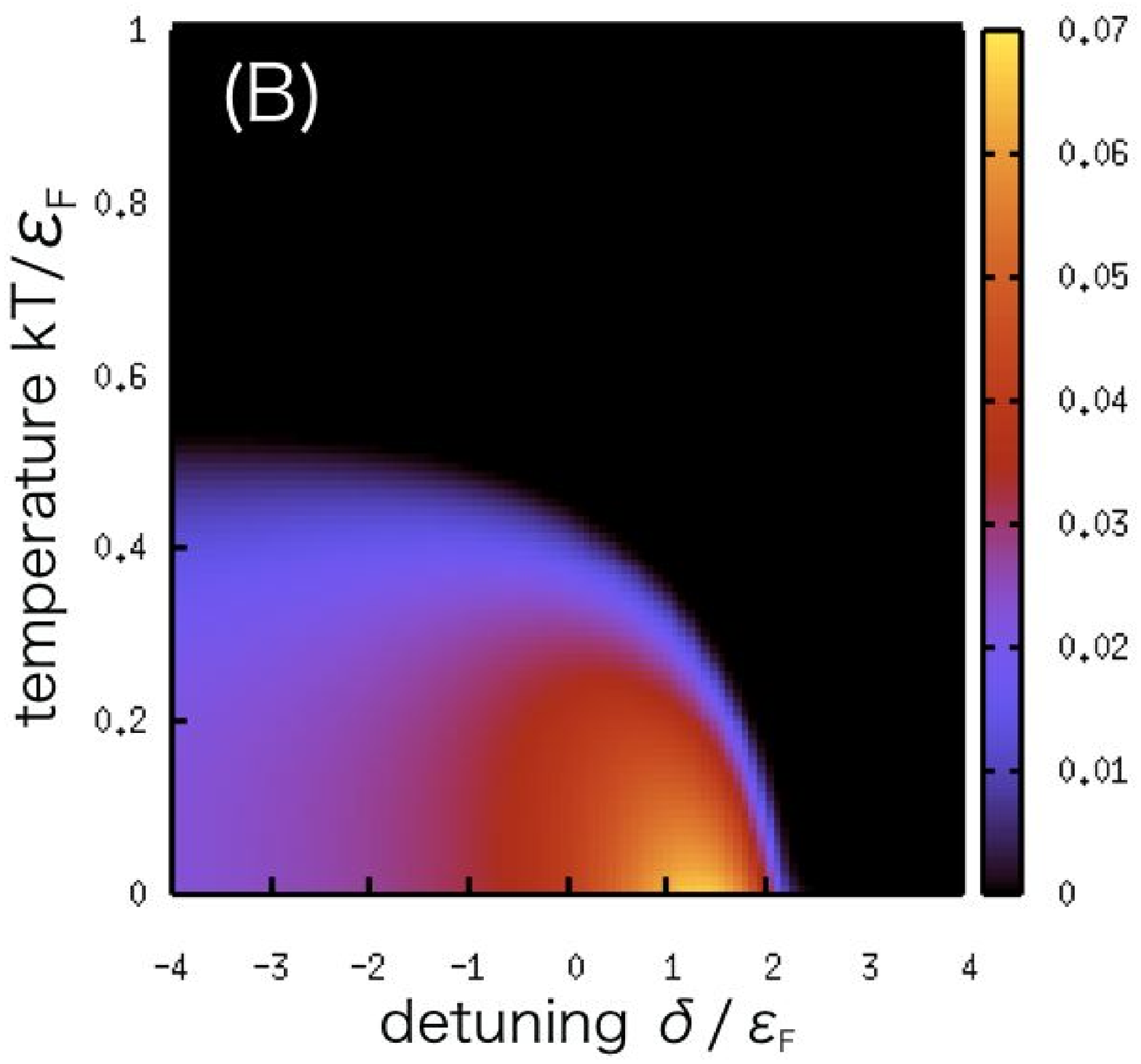}
\end{center}
\end{minipage}
\begin{minipage}{0.4\hsize}
\begin{center}
\includegraphics[width=6cm,height=6cm,keepaspectratio,clip]{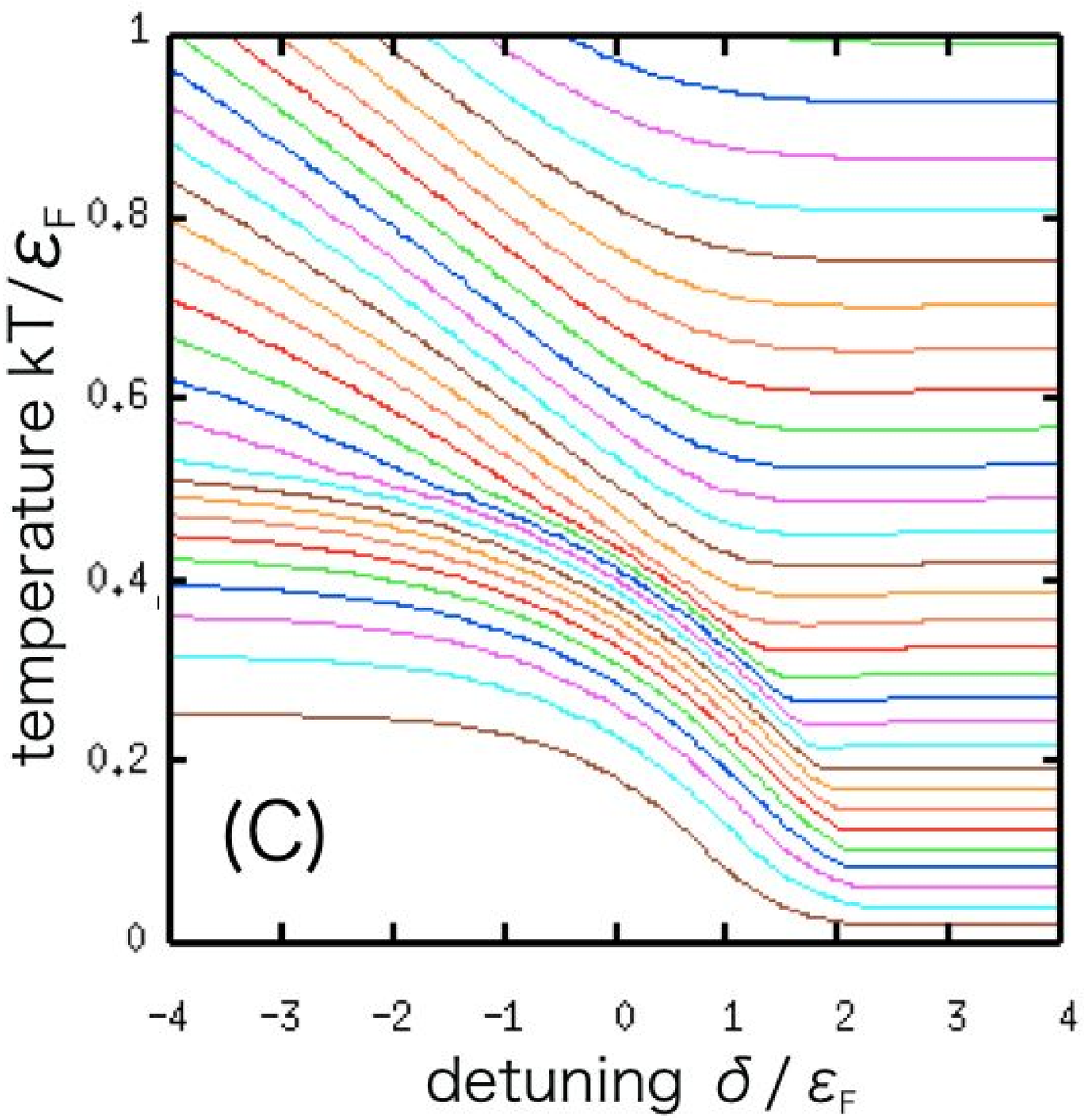}
\end{center}
\end{minipage}
\end{center}
\caption{
(Color online) 
(A) Conventional phase diagram of 
the condensed molecular fraction $\eta_{mc}$. 
(B) Conventional phase diagram of 
the condensed Cooper pair fraction $\eta_{p}$. 
Note the change in colorscale between panels (A) and (B). 
Each fraction in the conventional phase diagram is plotted against 
the temperature $T$  and the detuning $\delta$, 
normalized by the Fermi energy. 
(C) Contours of constant entropy. The paths 
are plotted against the temperature $T$ 
and the detuning $\delta$, 
as the detuning is varied adiabatically. 
We set the coupling strength $\alpha = 0.4 \varepsilon_{\rm F}$.
}
\label{LDATS.fig}
\end{figure}

Before showing contours of constant entropy, 
we plot the conventional phase diagrams of 
the condensed molecular fraction $\eta_{mc} = 2N_{mc}/N_{\rm tot}$ 
and the condensed Cooper pair fraction 
$\eta_{p} = 2N_{p}/N_{\rm tot}$ separately 
against the temperature $T$ and the detuning 
$\delta$ in Figs.~\ref{LDATS.fig} (A) and~\ref{LDATS.fig} (B). 
From Fig.~\ref{LDATS.fig} (B), 
we find that 
the number of Cooper pairs at low temperatures is 
peaked above the resonance 
in the BCS regime, 
while its contribution to the total  condensed fraction $\eta_{c}$ 
is very small. 
This is because we deal 
with the weak coupling (narrow Feshbach resonance). 
In the case of the strong coupling (broad Feshbach resonance), 
the Cooper pair contribution in the fermionic condensate fraction becomes dominant, and the single channel model describes effectively such systems in the vicinity of the resonance~\cite{ohashi2}. 
However, as shown in ref. 23, 
the conventional phase diagrams are almost the same for both cases, 
when plotted against $(k_{\rm F}a_{s})^{-1}$ instead of $\delta$ 
($k_{\rm F}$ is the Fermi wavenumber and $a_{s}$ is the $s$-wave scattering length). 
ref. 23 
also shows that the behaviors 
of the total number of 
bosons (including both Cooper pairs in the open channel and 
bare molecules of the closed channel) are almost the same for the two 
cases. 
We thus expect that the adiabatic phase diagram discussed below 
also describes the qualitative behavior in the broad resonance case.

In Fig.~\ref{LDATS.fig} (C), we show contours of 
constant entropy 
in the $\delta$ - $T$ plane. 
We recognize those lines, as the paths 
that are traversed 
as the detuning $\delta$ 
is swept adiabatically, 
starting from the right side 
of the resonance 
at a large value of the detuning $\delta$. 
We see that the temperature increases
 as the detuning $\delta$ 
is lowered adiabatically. 
This temperature increase 
on the path of constant entropy 
can be understood as follows: 
first, due to the conversion of pairs of atoms into molecules, 
the system loses degrees of freedom; 
second, the condensed molecules 
does not contribute to the entropy; third, the existence of Cooper pairs 
also suppresses the entropy. 
For these reasons, 
the system must then heat up in order to conserve the entropy.

The relation between the initial temperature 
of the atomic gas 
and the final temperature of the molecular gas 
has been given for 
the ideal gas mixture of fermionic atoms and 
bosonic molecules by Williams {\it et al}.~\cite{williams}. 
In the low-temperature limit $T/T_{\rm F} \ll 1$, 
one finds 
\begin{equation}
\frac{T_{f}}{T_{\rm F}} = \left[ \frac{\pi^{2}}{12\zeta (4)}\right]^{1/3}
\left(\frac{T_{i}}{T_{\rm F}}\right)^{1/3}, 
\end{equation}
while in the high-temperature limit, $T/T_{\rm F} \gg 1$, 
the relation becomes 
\begin{equation}
\frac{T_{f}}{T_{\rm F}} = 6^{1/3} {\rm e}^{4/3} \left(\frac{T_{i}}{T_{\rm F}}\right)^{2}.
\end{equation}
The above relations can be derived 
by connecting the initial entropy 
of a pure atomic gas 
and the final entropy of a pure molecular gas. 
Similar expressions have been derived by 
Carr {\it et al}.~\cite{carr} in connection with 
the idea of cooling a Fermi gas by adiabatically 
increasing the magnetic field starting from a molecular gas. 

\begin{figure}
\begin{center}
\includegraphics[width=7cm,height=7cm,keepaspectratio,clip]{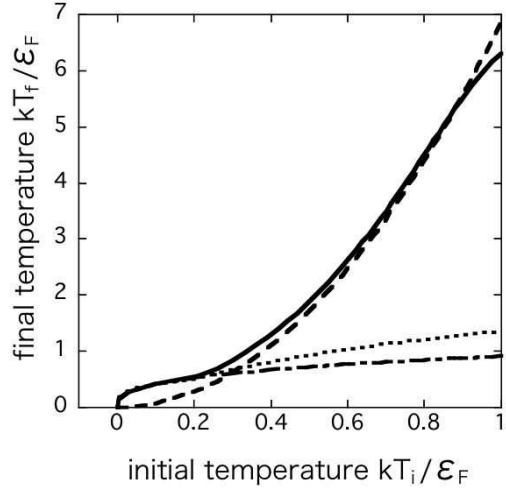}
\end{center}
\caption{
The final temperature versus the initial temperature 
where the final detuning is 
$\delta/\varepsilon_{\rm F} = -100$ (solid line) 
and 
$\delta/\varepsilon_{\rm F} = -4$ (dotted line). 
The dashed and dot-dashed lines 
represent the high and low temperature limit.  
We set the coupling strength $\alpha = 0.4 \varepsilon_{\rm F}$.
}
\label{LDATfTi.fig}
\end{figure}

In Fig.~\ref{LDATfTi.fig}, we plot the final temperature 
in a deeply BEC region and a moderately BEC region 
as a function of the initial temperature. 
The solid line is obtained by a numerical calculation 
in a deeply BEC region 
where the detuning is 
$\delta = -100\varepsilon_{\rm F}$. 
The dotted line is obtained by a numerical calculation 
in a moderately BEC region 
where the detuning is 
$\delta = -4\varepsilon_{\rm F}$. 
The dashed line represents the high-temperature limit 
for an ideal gas, 
while the dot-dashed line represents the low-temperature limit 
for an ideal gas. 
The low-temperature approximation 
agrees with the both of numerical solutions. 
The high temperature approximation does not agree with 
the numerical solution of a moderately BEC region, 
because atoms appear 
when the temperature is not sufficiently high 
and the concept of connecting 
the entropy of a pure atomic gas and 
that of a gas consisting entirely of molecules 
breaks down. 
On the other hand, 
the high temperature approximation 
agrees with a numerical solution 
of a deeply BEC region 
except for the high initial temperature regime 
where some atoms remain at the end of the sweep. 
These agreements are reasonable 
since our model assumes 
the narrow resonance case, leaving out the Cooper pairs 
outside the condensate. 
The system is equivalent to the ideal gas mixture 
for temperatures above $T_{c}$. 

\begin{figure}[htbp]
\begin{center}
\begin{minipage}{0.4\hsize}
\begin{center}
\includegraphics[width=7cm,height=7cm,keepaspectratio,clip]{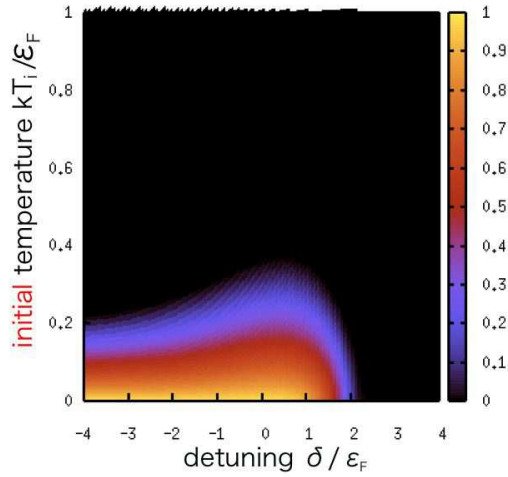}
\end{center}
\end{minipage}
\end{center}
\caption{(Color online) 
The adiabatic phase diagram. 
The total fermionic condensate fraction 
$\eta_{c}=\eta_{mc}+\eta_{p}$ 
is plotted against the initial temperature $T_{\rm i}$ and the detuning 
$\delta$. 
}
\label{LDATiNc.fig}
\end{figure}

In Fig.~\ref{LDATiNc.fig},  
we plot the fermionic condensate fraction 
$\eta_{c} = 2N_{c}/N_{\rm tot}$ 
against the detuning $\delta$ 
and the initial temperature of the atomic Fermi gas measured 
before a sweep of the magnetic field, 
as in the experimental phase diagrams 
of refs. 1 and 3. 
The initial temperature $T_{i}$ is found 
by equating the total entropy to the initial entropy 
$S_{\rm tot}(\delta, T) = S_{i}(T_{i})$. 

In Figs.~\ref{LDATS.fig} (A) and~\ref{LDATS.fig} (B), 
the transition temperature is 
$T_{c} \simeq 0.53T_{\rm F}$ in the BEC region 
where the detuning 
is $\delta = -4 \varepsilon_{\rm F}$. 
On the other hand, in Fig.~\ref{LDATiNc.fig}, 
the transition temperature in terms of 
the initial temerature is 
$T_{i,c} \simeq 0.21T_{\rm F}$ at $\delta = -4 \varepsilon_{\rm F}$. 
This behavior is consistent with Williams {\it et al}.~\cite{williams} 
for an ideal gas model, 
where the transition temperature is $T_{c} \simeq 0.51T_{\rm F}$  
and 
the transition temperature in terms of 
the initial temerature is 
$T_{i,c} \simeq 0.21T_{\rm F}$ 
at $\delta = -4 \varepsilon_{\rm F}$. 
The difference in the transition temperature 
between our model and the ideal gas model 
becomes smaller 
when we translate 
the conventional transition temperature 
to the transition temperature 
in terms of the initial temperature. 
This translated transition temperature $T_{i,c}$ 
in the BEC limit 
is comparable with transition temperature 
observed in the experiments~\cite{regal, zwierlein}. 
Here we comment on the finite energy gap of bosonic excitations 
discussed in {\S} 2. 
Because of this artificial gap, our mean-field results should be 
unreliable in the low temperature region 
where $k_{B}T < \delta - 2\mu$. 
However in our adiabatic phase diagram, 
for $\delta < 0.5\varepsilon_{\rm F}$, 
the real temperature satisfies the inequality 
$k_{B}T \gg \delta - 2\mu$ 
as long as the initial temperature satisfies 
$k_{B}T _{i} > 0.05\varepsilon_{\rm F}$. 
The region where the real temperature satisfies 
$k_{B}T \gg \delta - 2\mu$ becomes wider 
as one goes into the BEC side of the resonance. 
On the other hand, 
for $\delta > 0.5\varepsilon_{\rm F}$, 
the total number of molecules is so small at low temperatures, 
that the bosonic excitations make no significant contribution 
to the thermodynamic. 
Therefore, the finite energy gap in the bosonic excitation spectrum 
due to the mean-field treatment does not affect 
our adiabatic phase diagram significantly except 
for the low temperature region 
$k_{B}T_{i} < 0.05\varepsilon_{\rm F}$ in the crossover region 
$\delta \sim 0$.

\begin{figure}
\begin{center}
\includegraphics[width=7cm,height=7cm,keepaspectratio,clip]{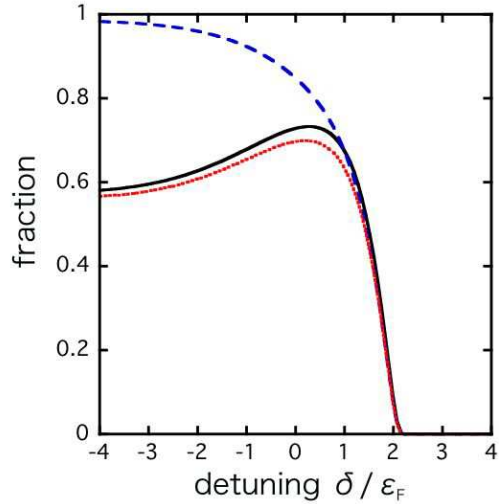}
\end{center}
\caption{(Color online) 
The fermionic condensate fraction $\eta_{c}$, 
the condensed molecular fraction $\eta_{mc}$ and 
the molecular fraction $\eta_{m}$ 
for the initial temperature $T_{\rm i} = 0.08T_{\rm F}$ are 
plotted against the detuning 
$\delta$.  
The solid line represents 
the fermionic condensate fraction $\eta_{c}$. 
The dotted line represents the condensed molecular fraction $\eta_{mc}$. 
The dashed line represents the molecular fraction $\eta_{m}$. 
We set the coupling strength $\alpha = 0.4 \varepsilon_{\rm F}$.
}
\label{LDATi008.fig}
\end{figure}

In Fig.~\ref{LDATi008.fig}, 
we plot the fermionic condensate, the condensed molecular, and the total molecular  fractions with an initial temperature $T_{i}/T_{\rm F} = 0.08$ 
as a function of the detuning.  
As the detuning is lowered, 
the fermionic condensate fraction increases. 
Subsequently it decreases after having a peak near the resonance. 
This behavior is consistent with experiments~\cite{regal, zwierlein}. 
Williams {\it et al}.~\cite{williams} finds a similar result 
for an ideal gas mixture, 
though they normalize 
the condensate fraction by the number of molecules.

\section{Molecular Conversion Efficiency}
Hodby {\it et al}.~\cite{hodby} 
have investigated 
the molecular conversion efficiency 
in bosonic $^{85}$Rb and fermionic $^{40}$K. 
They found that the molecular conversion efficiency 
does not reach 100\% even for an adiabatic sweep, 
but saturates at a value, that depends on initial peak phase-space density. 
Their experimental data is simulated 
by a Monte Carlo-type simulation~\cite{hodby} 
based on a model assuming that 
pairing formation occurs 
when two atoms are close enough in the phase space. 

On the other hand, 
Williams  {\it et al}.~\cite{williams-conversion} 
discussed the molecular conversion efficiency assuming 
the molecular formation rate vanishes 
when the detuning $\delta$ is negative. 
Their result is derived by 
using the coupled atom-molecule Boltzmann equations. 
They calculated 
the conversion efficiency 
from the molecular fraction at 
the detuning $\delta = 0$ 
as a function of the initial peak phase space density, 
and found good agreement with 
the experimental data although they treated 
an ideal gas mixture 
of Fermi atoms and Bose molecules. 
We note that the initial temperature and the initial phase space density  have 
a one-to-one correspondence. 

\begin{figure}[htbp]
\begin{center}
\begin{minipage}{0.4\hsize}
\begin{center}
\includegraphics[width=7cm,height=7cm,keepaspectratio,clip]{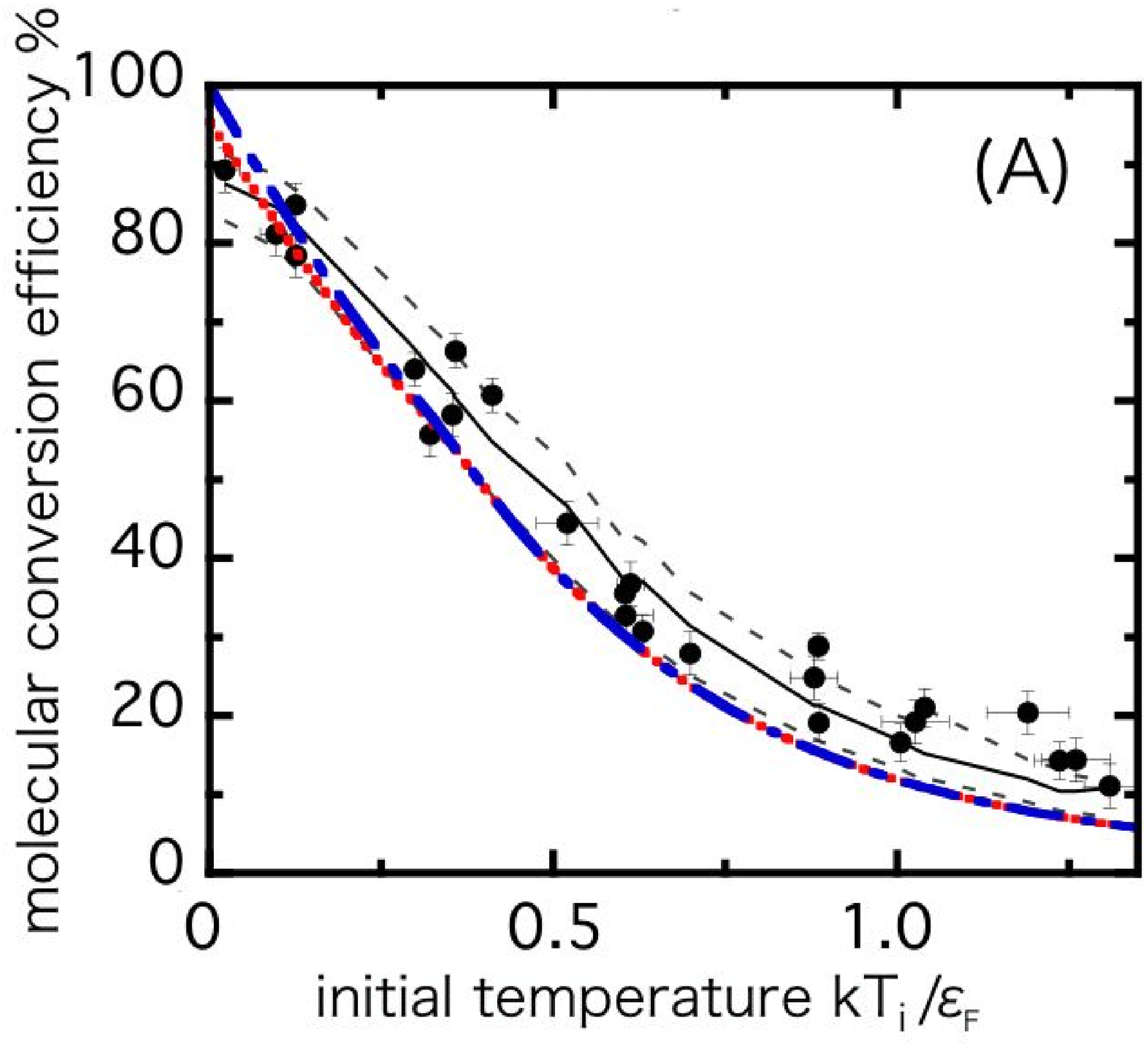}
\end{center}
\end{minipage}
\begin{minipage}{0.05\hsize}
\begin{center}
\end{center}
\end{minipage}
\begin{minipage}{0.4\hsize}
\begin{center}
\includegraphics[width=6.5cm,height=6.5cm,keepaspectratio,clip]{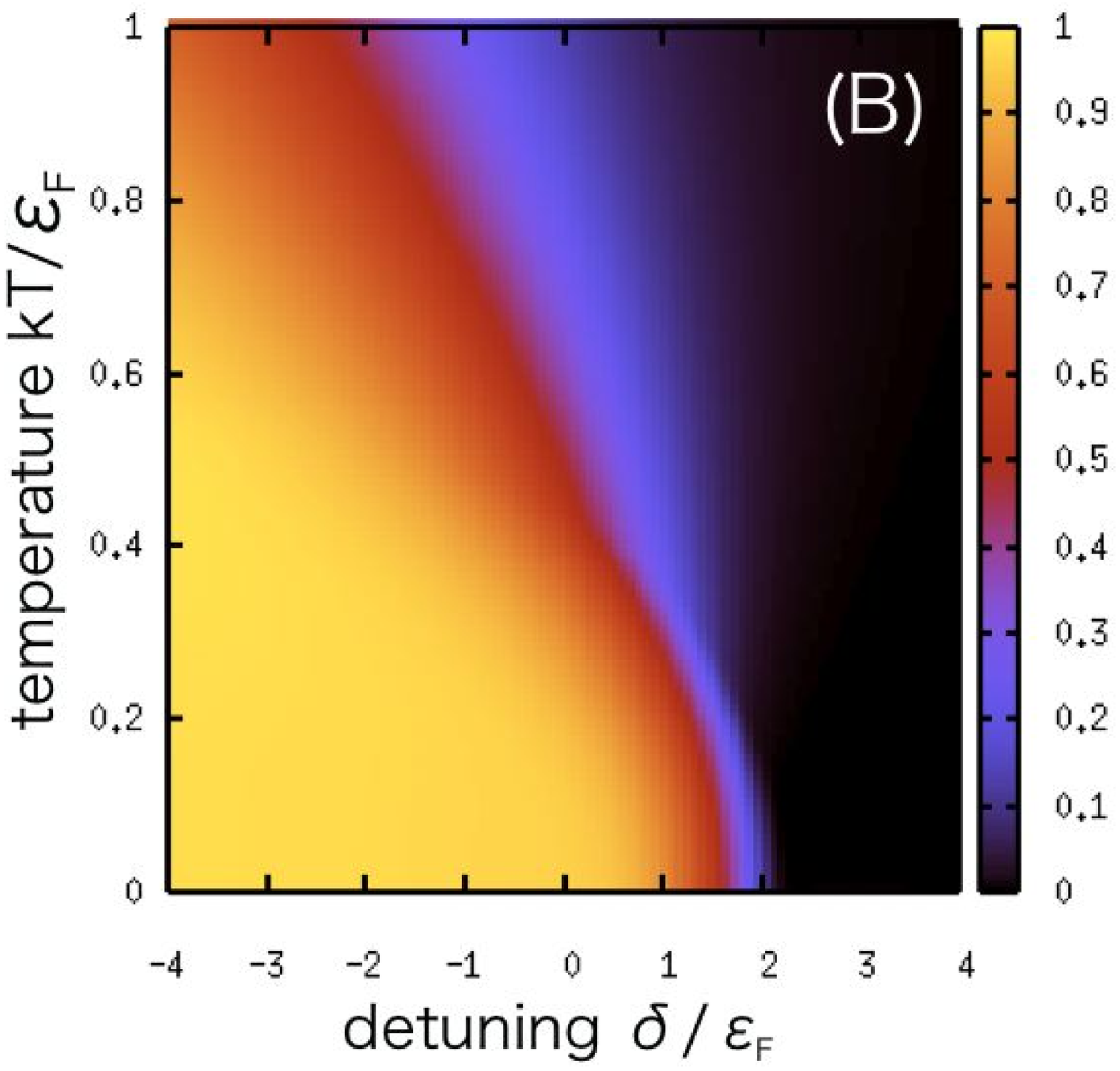}
\end{center}
\end{minipage}
\end{center}
\caption{(Color online) 
(A) The molecular conversion efficiencies 
are plotted against the initial temperature. 
The dotted line represents the our model, 
while the dot-dashed line is the result 
for an ideal gas mixture~\cite{williams-conversion}. 
The dots represent 
the experimental data of Hodby {\it et al}.~\cite{hodby}. 
The solid line represents their simulation result, and 
the dashed lines give the uncertainty 
of the pairing parameter in their simulation. 
(B) For reference, the molecular fraction $\eta_{m}$ is plotted 
as a function of temperature and the detuning. 
The fraction at the resonance position $\delta=0$ 
and a temperature adiabatically connected with the initial temperature 
are 
used for the conversion efficiency in Fig. (A). 
We set the coupling strength $\alpha = 0.4 \varepsilon_{\rm F}$.
}
\label{LDAMCE.fig}
\end{figure}

In the experiment for $^{40}$K, 
the conversion fraction does not depend on the inverse sweep rate 
within a range from $640$ to $2900 \mu s/G$. 
In this range, therefore the conversion process can be considered as 
adiabatic~\cite{hodby}. 
According to the kinetic theory of ref. 18, 
in this regime the chemical equilibration between atoms and 
molecules also brings about the thermal equilibration. 
Assuming this adiabatic limit 
and that the principle controlling the maximum molecular conversion efficiency proposed by 
Williams {\it et al}. also applies when including 
the resonant interaction, 
we calculate the molecular conversion efficiency 
using $\eta_{m}(\delta = 0)$ 
as a function of initial temperature, 
as shown in Fig.~\ref{LDAMCE.fig} (A), 
where $\eta_{m}$ is the molecular fraction $2N_{m}/N_{\rm tot}$. 
We note that the Feshbach resonance of $^{40}$K is 
known to be broad~\cite{hodby}. 
Although our mean-field theory is 
only applicable to the narrow resonance case, 
we can see the qualitative effect of 
the resonance interaction from Fig.~\ref{LDAMCE.fig} (A). 
The dotted line represents the result from our model, 
while the dot-dashed line represents the ideal gas mixture 
of Fermi atoms and Bose molecules~\cite{williams-conversion}. 
The dots represent 
the data of Hodby  {\it et al}.~\cite{hodby},  
and the solid line represents their simulation result. 
Dashed lines represent the uncertainty 
of the pairing parameter 
in their simulation. 
As noted above, the weak coupling constant we used does not 
directly correspond to the experiment, 
and thus the comparison between the theory 
and experiment is only qualitative. 
Nevertheless we find that 
our model agrees well with 
the trend of experimental data. 
The transition temperature $T_{c}$ at $\delta=0$ 
is $0.45T_{\rm F}$, 
which corresponds to 
the transition temperature in terms of 
initial temperature 
$T_{i,c} = 0.36T_{\rm F}$. 
Above $T_{i,c}$ 
our model is identical to the ideal gas mixture case,  since
we are treating the resonant interaction 
in a mean-field theory. 
For reference, we show the molecular fraction $\eta_{m}$ 
plotted as a function of 
temperature and the detuning in Fig.~\ref{LDAMCE.fig} (B). 
The fraction at the resonance position $\delta = 0$ and 
a temperature adiabatically connected with the initial temperature are 
used for the conversion efficiency in Fig.~\ref{LDAMCE.fig} (A).

The resonant interaction is more important 
for lower initial temperatures. 
At the initial temperature $T_{i} = 0$, 
the molecular conversion efficiency is 100\% 
in the ideal gas mixture model. 
On the other hand, 
the experimental data at $T_{i}\rightarrow0$ 
is suppressed from 100\%. 
The efficiency in our model at $T_{i}=0$ 
is also less than 100\%, 
because the resonant interaction suppresses the molecular conversion. 
We note that the resonance position may shift from the ideal gas case 
due to the atom-molecule interaction, 
which may also shift the result for the conversion efficiency  quantitatively. 
However, the qualitative relation between the molecule 
conversion efficiency and the initial temperature 
will remain unchanged.

\section{Conclusion} 
In this paper, we have calculated the phase diagrams 
for resonantly-coupled system of Fermi atoms and Bose molecules. 
We determined the paths of constant entropy traversed 
in the phase diagram 
as the detuning is lowered adiabatically. 
The adiabatic phase diagram of the fermionic condensate fraction 
composed of the condensed molecules and the Cooper pairs 
was plotted against the detuning 
$\delta$ and the initial temperature 
of a pure atomic gas. 
The adiabatic phase diagram 
allowed us to compare the theory with experiments, 
since in the experients~\cite{regal, zwierlein}, 
the condensate fraction 
have been measured against the initial temperature and the magnetic field. 
We obtained the transition temperature 
in terms of the initial temperature 
$T_{i,c} \simeq 0.21T_{\rm F}$ in the BEC regime. 
Finally, the molecular conversion efficiency 
was plotted as a function of the initial temperature. 
We found that the resonant interaction suppresses 
the complete conversion.

The present work is an extension of the recent works 
by Williams {\it et al}.~\cite{williams, williams-conversion} 
to include the resonant 
interaction using a mean-field theory. 
Although our model assumes a narrow resonance, 
the mean-field calculations help qualitative understanding 
of the recent experiments on the BCS-BEC crossover. 
In particular, the suppression of $T_{i,c}$ in the BEC limit and 
the behavior of the molecular conversion efficiency will be 
qualitatively unchanged even in the broad resonance case. 
In order to compare 
the theory with experiments in a quantitative fashion, 
we must include the Cooper pairs outside the condensate, 
which are neglected in our mean-field calculation. 
We must also express the resonance position in terms of the magnetic field 
rather than the detuning. 

\section{Acknowledgements}
We thank Chikara Ishii and Satoru Konabe 
for helpful discussions and continuous encouragements. 
We also thank  Allan Griffin for useful comments.

\end{document}